\documentclass[12pt]{article}
\usepackage{epsfig}
\usepackage[cp1251]{inputenc}
\usepackage[english]{babel}

\newcommand{\fr}[2]{{\displaystyle \frac{#1}{#2}}}

\newcommand{\rf}{{{r}{\!}{f}}}

\title{DYNAMICS OF RAREFACTION WAVES IN GENERAL-RELATIVISTIC COLLAPSING CLOUDS}
\author
{ Zhilkin A.G.\thanks{\small Submitted in {\it Gravitation \& Cosmology}}\\ %
\textit{\small Institute of Astronomy RAS, Russia, e-mail: zhilkin@inasan.ru}\\%
}
\date{}

\begin{document}
\maketitle

\begin{abstract}
Formation and evolution of general relativistic collapse
inhomogeneity is investigated. It is shown that a rarefaction wave
forms at initial collapse stages and propagates inside from the cloud
boundary to its center. The focusing time of the rarefaction wave is
found. In massive clouds, the rarefaction wave focusing time equals
the free-fall time. The collapse of such clouds leads to black hole
formation. In low-mass clouds, the rarefaction wave focusing time is
less than the free-fall time. After focusing, the collapse of such
clouds becomes fully inhomogeneous and can be sufficiently
decelerated by the pressure gradient.
\end{abstract}

\section{Introduction}

The study of general relativistic gravitational collapse is a complex
problem because of mathematical difficulties appearing when solving
the corresponding non-linear partial differential equations even in
spherical symmetry \cite{Misner1964}. Only a few exact solutions are
known for any special cases \cite{Oppenheimer1939}. More realistic
models for gravitational collapse can be investigated using numerical
simulations (see, e.g., \cite{Shapiro1980}).

An initially homologous collapsing cloud with zero pressure (dust
cloud) remains homologous during the collapse. But contraction of an
initially homologous cloud with non-zero pressure develops with time
in a sufficiently non-homologous way because a pressure gradient must
necessarily develop at the cloud surface \cite{Harada2001}. Even if
the cloud is initially surrounded by an ambient gas of the same
pressure as the cloud, the further contraction breaks this
equilibrium \cite{Larson1972}. A rarefaction wave propagates in from
the cloud surface to its center with sound speed. The rarefaction
front divides the cloud into two parts \cite{Zeldovich1970,
Dudorov2003, Harada2001}. The pressure gradient becomes zero in the
internal region. However, some inhomogeneity forms in the external
region. This dynamical inhomogeneity can evolve in a self-similar
manner at last stages of contraction \cite{Ori1990, Cai2005}.

The dynamics of rarefaction wave has been studied for
non-relativistic (protostellar) collapse of an isothermal cloud
\cite{Truelove1998}. The effects of rotation \cite{Tsuribe1999} and
magnetic field \cite{Dudorov2003} have been taken into account. In
this paper, a general-relativistic analogue of this problem is
considered under the condition of spherical symmetry.

In the next section, the basic equations are described. In Sec. 3,
the flow structure in the inner homogeneously collapsing core is
considered. The equation of rarefaction front propagation is solved
in Sec. 4. In Sec. 5, the radial null geodesics are investigated.
Finally, Sec. 6 presents the main results and conclusions.

\section{Basic equations}

Let us work in a system of spherical coordinates ($t$, $R$, $\theta$,
$\varphi$) moving at each point with the gas located at that point
(comoving or Lagrangian coordinates). The interior space-time
geometry for a collapsing cloud in spherical symmetry can be
described by the metric of a ``collapsing star'' (see, e.g.,
\cite{Misner1973}):
\begin{equation}\label{eq201}
  ds^2 = e^{2\Phi} c^2 dt^2 - e^{2\Lambda}dR^2 - r^2 d\Omega^2,
\end{equation}
where $d\Omega^2=d\theta^2+\sin^2\theta d\varphi^2$. Here $\Phi$,
$\Lambda$, $r$ are some functions of the radial coordinate $R$ and
the time $t$ to be determined by the Einstein equations. Note that
the form of the line element (\ref{eq201}) is invariant under the
transformation: $R'=R'(R)$, $t'=t'(t)$.

In the comoving reference frame, the components of the four-velocity
of gas are $u^0 = e^{-\Phi}$, $u^1 = u^2 = u^3 = 0$. Therefore, for a
perfect fluid, the stress-energy tensor has following non-zero
components: $T^0_0 = e$, $T^1_1 = T^2_2 = T^3_3 = -P$, where $e$ is
the gas energy density (gas energy per unit proper rest volume) and
$P$ is the gas pressure.

Using these relations, we can obtain the following Einstein field
equations:
\begin{equation}\label{eq205}
  \fr{8\pi G}{c^4} P =
  e^{-2\Lambda}
  \left( \fr{{r'}^2}{r^2} + 2\Phi' \fr{r'}{r} \right) -
  \fr{1}{c^2} e^{-2\Phi}
  \left(
    2\fr{\ddot{r}}{r} + \fr{\dot{r}^2}{r^2} -
    2\dot{\Phi} \fr{\dot{r}}{r}
   \right) -
   \fr{1}{r^2},
\end{equation}
\begin{equation}\label{eq207}
  \fr{8\pi G}{c^4} e =
  -e^{-2\Lambda}
  \left(
   2\fr{r''}{r} + \fr{{r'}^2}{r^2} -
   2 \Lambda' \fr{r'}{r}
  \right) +
  \fr{2}{c^2} e^{-2\Phi}
  \left(
    \dot{\Lambda} \fr{\dot{r}}{r} +
    \fr{1}{2} \fr{\dot{r}^2}{r^2}
   \right) + \fr{1}{r^2},
\end{equation}
\begin{equation}\label{eq208}
  0 = \fr{\dot{r}'}{r} -
  \dot{\Lambda} \fr{r'}{r} -
  \Phi' \fr{\dot{r}}{r},
\end{equation}
where the prime means a derivative with respect to the radial
coordinate $R$ and the overdot means a derivative with respect to the
time $t$. The equations of gas motion give
\begin{equation}\label{eq209}
  \fr{\dot{e}}{e + P} =
  -\dot{\Lambda} - 2 \fr{\dot{r}}{r}, \,\,
  \fr{P'}{e + P} = - \Phi'.
\end{equation}

The system (\ref{eq205}-\ref{eq209}) is still not complete. We must
add any equation of state to close this system. In this paper, we
assume the following equation of state:
\begin{equation}\label{eq211}
  P = (\kappa - 1) e,
\end{equation}
where $\kappa=\mbox{\rm const}$ ($1 \le \kappa \le 2$). From this
equation it follows that the sound speed in the gas is $a =
\sqrt{\kappa - 1} c$.

Eq. (\ref{eq211}) can correspond to two reasonable astrophysical
situations \cite{Ori1990}. In the first case ({\it protocluster} and
{\it protogalactic} clouds) this equation can be considered as the
equation of state of a non-relativistic isothermal ideal gas when
adiabatic heating and cooling is neglected and therefore the internal
energy is negligible compared to the rest energy. In the second case
({\it presupernova cores}), Eq. (\ref{eq211}) can be considered as an
extreme relativistic limit of an adiabatic (or polytropic) equation
of state when the rest energy density is negligible relative to the
internal energy.

\section{Homogeneously collapsing core}

Inside the internal homogeneous core of a collapsing cloud, the
pressure gradient is zero. From the second equation in (\ref{eq209})
it follows that $\Phi' = 0$ and therefore $\Phi=\Phi(t)$. Thus we can
define another time variable $T=T(t)$ such that $\Phi=0$. It is clear
that the time $T$ coincides with the proper time of the gas. In
addition, suppose that the radial coordinate $R$ satisfies the
relation $r(T, R) = R (1 - \eta(T))$, where $\eta(T)$ is a function
of the proper time $T$ such that $0 \le \eta \le 1$. The value
$\eta=1$ corresponds to free-fall time (without the pressure
gradient) for the considered homogeneous region.

From Eq. (\ref{eq208}) we get the following solution:
\begin{equation}\label{eq302}
  \Lambda =
  \ln \fr{1 - \eta}{\sqrt{1 + f(R)}},
\end{equation}
where $f(R)$ is some function. Combining (\ref{eq302}) and
(\ref{eq209}), we find
\begin{equation}\label{eq303}
  e = \fr{e_0}{(1 - \eta)^{3\kappa}},
\end{equation}
where $e_0$ is initial energy density.

Now we can use Eqs. (\ref{eq302}), (\ref{eq303}) to transform Eq.
(\ref{eq207}) to the following form:
\begin{equation}\label{eq304}
  \fr{8\pi G}{c^4} e =
  \fr{3}{c^2} \fr{\dot{\eta}^2}{(1 - \eta)^2} -
  \fr{f}{R^2 (1 - \eta)^2} -
  \fr{f'}{R (1 - \eta)^2}.
\end{equation}
Since the left-hand side of this equation is a function of the proper
time $T$, we see that $f = f_0 R^2$, where $f_0$ is some constant.
Taking into account the initial values $\eta(0)=1$ and
$\dot{\eta}(0)=0$, we can find: $f_0 = -8\pi G e_0/(3c^4)$.

The invariant line element (\ref{eq201}) can be written as
\begin{equation}\label{eq306}
  ds^2 =
  c^2 dT^2 -
  (1-\eta)^2
  \left(
   \fr{dR^2}{1 - \fr{8\pi G e_0}{3c^2}R^2} + R^2 d\Omega^2
  \right).
\end{equation}
Note that this metric corresponds to the well-known Friedmann
solution for a closed homogeneous isotropic universe with uniform
pressure.

Now we can combine the expression for energy density (\ref{eq303})
with (\ref{eq304}) to obtain an equation for the function $\eta(T)$:
\begin{equation}\label{eq307}
  \fr{d\eta}{dT} =
  \sqrt{\fr{8\pi G e_0}{3 c^2}}
  \left[
    \fr{1}{(1 - \eta)^{3\kappa - 2}} - 1
  \right]^{1/2}.
\end{equation}
This equation can be solved analytically in the special case
$\kappa=4/3$. For an arbitrary parameter $\kappa$, Eq. (\ref{eq307})
can be solved numerically.

\section{Propagation of a rarefaction wave}

The rarefaction wave front propagates in collapsing gas with a sound
speed. The location $R_{\rf}(T)$ of the rarefaction wave front
satisfies the equation:
\begin{equation}\label{eq401}
  e^{\Lambda} \fr{dR_{\rf}}{dT} = -a.
\end{equation}
After some manipulations, we can transform Eq. (\ref{eq401}) to the
form:
\begin{equation}\label{eq402}
  \fr{dr_{\rf}}{d\eta} =
  -\fr{\sqrt{\kappa - 1}
       (1 - \eta)^{\frac{3\kappa - 2}{2}}
       \sqrt{1 - \alpha^2 r_{\rf}^2}}
      {\alpha\sqrt{1 - (1 - \eta)^{3\kappa - 2}}}.
\end{equation}
where $r_{\rf}=R_{\rf}/R_0$, $R_0$ is the initial cloud radius,
\begin{equation}\label{eq403}
  \alpha = \sqrt{\fr{R_g}{R_0}}, \,
  R_g = \fr{2G M_0}{c^2} = \fr{8\pi G e_0 R_0^3}{3 c^4}.
\end{equation}

The parameter $\alpha$ controls the general relativity effect in the
problem. In the case of very small $\alpha$, the space-time curvature
is negligible, and the dynamics of the rarefaction wave can be
considered as non-relativistic. In the opposite limit, as $\alpha \to
1$, the problem is highly general-relativistic.

The solution of Eq. (\ref{eq402}) can be written as:
\begin{equation}\label{eq404}
  r_{\rf} = \fr{1}{\alpha}
  \sin \biggl(
   \arcsin\alpha +
   \fr{2\sqrt{\kappa - 1}}{3\kappa - 2}
   \arccos(1 - \eta)^{\frac{3\kappa - 2}{2}}
  \biggr).
\end{equation}
The solutions obtained, for the case $\kappa=5/3$ and various
$\alpha$, are shown in Fig. 1.

\begin{figure}[t]
\begin{center}
\epsfig{file=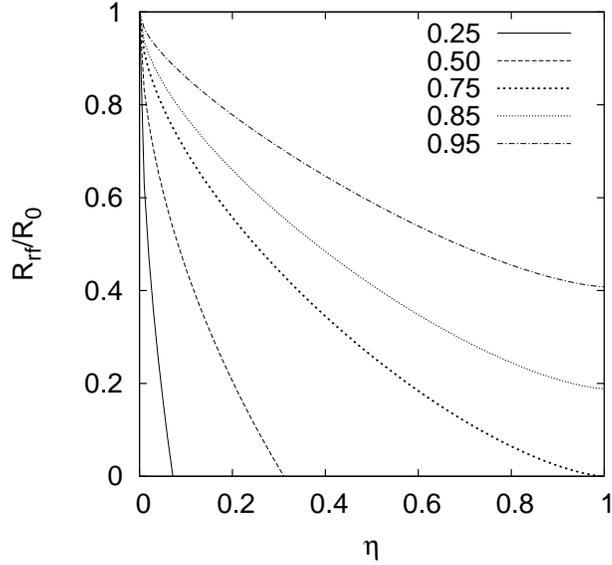,width=8cm}%
\caption{Functions $R_{\rf}(\eta)$ in the case $\kappa=5/3$ for
various values of parameter $\alpha$.}
\end{center}
\label{fg1}
\end{figure}

The rarefaction wave front reaches the center of the cloud at a
focusing time $T_{*}$. The value of $T_{*}$ can be determined by an
equivalent value $\eta_{*}=\eta(T_{*})$. There are two possible
regimes for propagation of rarefaction waves in general relativistic
collapsing clouds.

For sufficiently large $\alpha\ge\alpha_{*}$, where
\begin{equation}\label{eq405}
  \alpha_{*} = \sin \fr{\pi\sqrt{\kappa - 1}}{3\kappa - 2},
\end{equation}
we have $\eta_{*}=1$. This result means that the rarefaction wave
front reaches the cloud center at a moment of {\it big crunch} when
the central homogeneous core collapses to space-time singularity.

For relatively small values of parameter $\alpha<\alpha_{*}$ the
rarefaction wave front reaches the cloud center before the formation
of a central singularity. The focusing time is determined by the
value
\begin{equation}\label{eq406}
  \eta_{*} = 1 -
  \cos^{\frac{2}{3\kappa - 2}}
  \left(
    \fr{3\kappa - 2}{2\sqrt{\kappa - 1}}
    \arcsin\alpha
  \right).
\end{equation}
In such a cloud, after focusing of the rarefaction wave, a strong
pressure gradient can be formed. Hence the further collapse can be
sufficiently decelerated and can even pass to slower
quasi-hydrostatic contraction with formation of some stable object at
final stages of cloud evolution (e.g., a {\it neutron star}). Note
that, for a fixed initial radius $R_0$, a small value of the
parameter $\alpha$ corresponds to a small initial mass of the cloud.
The simplest expression for the focusing time can be obtained in
case $\kappa=4/3$:
\begin{equation}\label{eq407}
  T_{*} =
  \sqrt{\fr{3c^2}{8\pi G e_0}}
  \sin \left( \sqrt{3} \arcsin\alpha \right).
\end{equation}

The critical value $\alpha_{*}$ is a function of the parameter
$\kappa$. As $\kappa \to 0$ we have the asymptotic relation
$\alpha_{*} = \pi \sqrt{\kappa - 1}$. In particular, in the case of a
non-relativistic isothermal gas, the obtained criterion coincides
with the non-relativistic one (see \cite{Truelove1998, Tsuribe1999,
Dudorov2003}). The maximum value of $\alpha_{*}$ is reached at
$\kappa=4/3$. At this point, $\alpha_{*}=\sin{\pi\sqrt{3}/6}\approx
0.79$, and therefore the initial cloud radius is $R_0 \approx 1.61
R_g$.

\section{Null radial geodesics}

We can probe the space-time geometry in the internal homogeneous
region by emitting outgoing light rays. The equation for outgoing
radial null geodesics in the inner homogeneous region is
\begin{equation}\label{eq501}
  e^{\Lambda} \fr{dR}{dT} = c.
\end{equation}
After some simple manipulations we can transform this equation to
the following form:
\begin{equation}\label{eq502}
  \fr{dR}{d\eta} =
  \sqrt{\fr{3c^4}{8\pi G e_0}}
  \fr{(1 - \eta)^{\frac{3\kappa-4}{2}}
      \sqrt{1 - \fr{8\pi G e_0}{3c^4}R^2}}
     {\sqrt{1 - (1-\eta)^{3\kappa-2}}}.
\end{equation}
For null geodesics that are outgoing from the cloud center, we have
the following solution:
\begin{equation}\label{eq503}
  R =
  \fr{R_0}{\alpha}
   \sin \biggl\{
    \fr{1}{3\kappa-2}
     \left[
      \arcsin(1-\eta_0)^{\frac{3\kappa-2}{2}} -
      \arcsin(1-\eta)^{\frac{3\kappa-2}{2}}
     \right]
  \biggr\}.
\end{equation}
Here $\eta_0$ is an initial value of the parameter $\eta$ for a given
outgoing light ray.

\begin{figure}[t]
\begin{center}
\epsfig{file=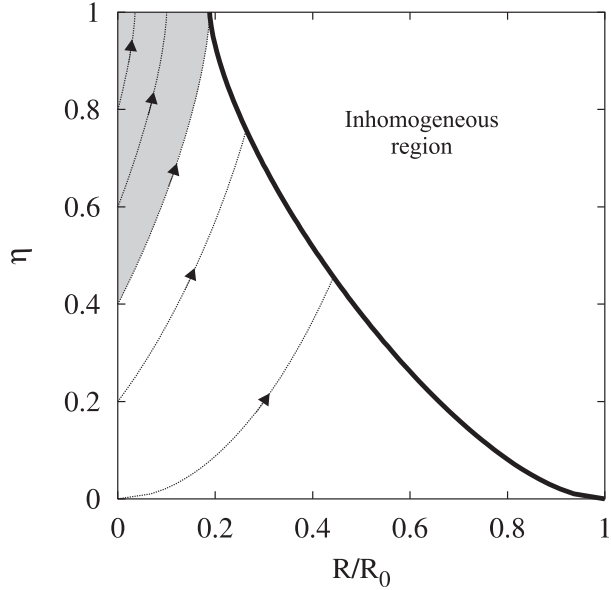, width=8cm}%
\caption{Space-time geometry of a collapsing cloud in case
$\kappa=5/3$ and $\alpha = 0.85$. Lines with arrows denote outgoing
radial light rays with various values of the initial parameter
$\eta_0$.}
\end{center}
\label{fg2}
\end{figure}

Fig. 2 shows the space-time geometry in a collapsing cloud with the
parameters $\kappa=5/3$ and $\alpha=0.85$ in terms of the coordinates
$\eta$ and $R$. The heavy solid line corresponds to the rarefaction
wave front. The arrowed lines denote outgoing radial light rays. Each
arrow points the direction of light ray propagation. Some outgoing
rays can reach the rarefaction wave front and then get out to the
outer inhomogeneous region of the collapsing cloud. Other rays cannot
reach the rarefaction wave front because they fall back to the cloud
center at $\eta=1$. Every light ray emitted at any point of the
shaded area cannot escape the internal homogeneous region. In other
words, we can conclude that an {\it event horizon} forms in the
considered cloud at final stages of its collapse.

Therefore, for clouds with $\alpha<\alpha_{*}$, there are no effects
that can retard the collapse, and some part of matter necessarily
falls under an event horizon with the formation of a {\it black
hole}. The obtained criterion for black hole formation in collapsing
clouds can be reformulated in another form: $M_0 > M_{*}$, where the
critical mass $M_{*} = 0.5 \alpha_{*}^2 c^2 R_0/G$ depends on the
initial cloud radius $R_0$. Note that in the case $\kappa \approx 1$
the value of the critical mass $M_{*}$ can be very small because in
such a cloud the critical parameter $\alpha_{*}$ is very close to
zero. On the other hand, an event horizon can appear in a collapsing
cloud after the rarefaction wave focusing. Hence the real critical
mass can even be less than $M_{*}$.

\section{Conclusion}

The basic results of this work are:\\

\noindent {\bf 1.} The problem of inhomogeneity of general
relativistic collapse of an initially homogeneous cloud surrounded by
an ambient gas with the same pressure is investigated. It is shown
that, at initial collapse stages, a rarefaction wave forms at the
cloud boundary and then propagates to the cloud center. The front of
this rarefaction wave divides the cloud into two parts. In the inner
homogeneous region, the space-time metric is like the Friedmann one.
In the outer region behind the rarefaction front, a strong
inhomogeneity forms.\\

\noindent {\bf 2.} The rarefaction wave front propagates in the
collapsing gas with the sound speed. The characteristic equation of
the rarefaction front propagation has been solved. The focusing time
of the rarefaction wave has been found.\\

\noindent {\bf 3.} An analysis reveals two possible regimes of
rarefaction wave dynamics. In very massive clouds, the rarefaction
wave focusing occurs at the central singularity formation time (big
crunch). The collapse of such clouds leads inevitably to black hole
formation. In a cloud with a relatively small mass, rarefaction wave
focusing takes place before the big crunch. After the rarefaction
wave focusing, the collapse of such a cloud can pass to quasi-static
contraction with further formation of some stable object.\\ \\

\textbf{Acknowledgement.} The author is grateful to Professor S.I.
Blinnikov for useful discussions. This work is supported by grants
RFBR (projects 05--02--17070, 05--02--16123) and RFBR-Ural (project
04--02--96050).

\end{document}